\documentclass[%
 reprint,
 amsmath,amssymb,
 aps,
]{revtex4-1}
\usepackage{amsthm}
\usepackage{graphicx}
\usepackage{float}
\usepackage{dcolumn}
\usepackage{bm}
\usepackage{extarrows}
\usepackage{hyperref}
\usepackage{multirow}
\usepackage{float}
\usepackage{color}
\usepackage{marvosym}
\usepackage{epstopdf}
\usepackage{subfigure}
\usepackage{array}
\usepackage{amssymb}
\usepackage{varioref}
\usepackage[utf8]{inputenc}
\usepackage{indentfirst}
\usepackage{subfigure}
\usepackage{amsmath}
\begin{document}

\preprint{APS/123-QED}

\title{Variational quantum algorithm-preserving feasible space for solving the uncapacitated facility location problem}

\author{Sha-Sha Wang $^{1}$}

\author{Hai-Ling Liu $^{1}$}
\author{Yong-Mei Li$^{1}$}
\author{Fei Gao $^{1}$}
\email{gaof@bupt.edu.cn}

\author{Su-Juan Qin $^{1}$}
\email{qsujuan@bupt.edu.cn}

\author{Qiao-Yan Wen $^{1}$}

\affiliation{$^{1}$ State Key Laboratory of Networking and Switching Technology, Beijing University of Posts and Telecommunications, Beijing 100876, China}
\date{\today}

\begin{abstract}
The Quantum Alternating Operator Ansatz (QAOA+) is one of the Variational Quantum Algorithm (VQA) specifically developed to tackle combinatorial optimization problems by exploring the feasible space in search of a target solution. For constrained optimization problems with unconstrained variables, which we call Unconstrained-Variables Problems (UVPs), the mixed operators in the QAOA+ circuit are applied to the constrained variables, while the single-qubit rotating gates $R_X$ operate on the unconstrained variables. The expressibility of this circuit is limited by the shortage of two-qubit gates and the parameter sharing in the $R_X$, which consequently impacts the performance of QAOA+ for solving UVPs. Therefore, it is crucial to develop a suitable ansatz for UVPs. In this paper, we propose the Variational Quantum Algorithm-Preserving Feasible Space (VQA-PFS) ansatz, exemplified by the Uncapacitated Facility Location Problem (UFLP), that applies mixed operators on constrained variables while employing Hardware-Efficient Ansatz (HEA) on unconstrained variables. The numerical results demonstrate that VQA-PFS significantly enhances the success probability and exhibits faster convergence compared to QAOA+, Quantum Approximation Optimization Algorithm (QAOA), and HEA. Furthermore, VQA-PFS reduces the circuit depth dramatically in comparison to QAOA+ and QAOA. Our algorithm is general and instructive in tackling UVPs.
\end{abstract}

\pacs{Valid PACS appear here}
\maketitle


\section{Introduction}
By harnessing quantum effects, quantum computers offer computational advantages over classical computers, delivering polynomial or even exponential speedups for specific problems, such as integer factorization \cite{1}, unstructured data search \cite{2}, linear regression \cite{4,5}, quantum error correction \cite {3}, matrix computation \cite{11,12,13,14}, and cryptanalysis \cite{16}. However, the current quantum hardware devices only support a limited number of physical qubits and limited gate fidelity, which makes the above quantum algorithms unable to be implemented on near-term devices.

\begin{figure*}[htbp]
	\centering
	\begin{minipage}{0.4\linewidth}
		\centering
		\label{fig:11(a)}\includegraphics[width=\linewidth]{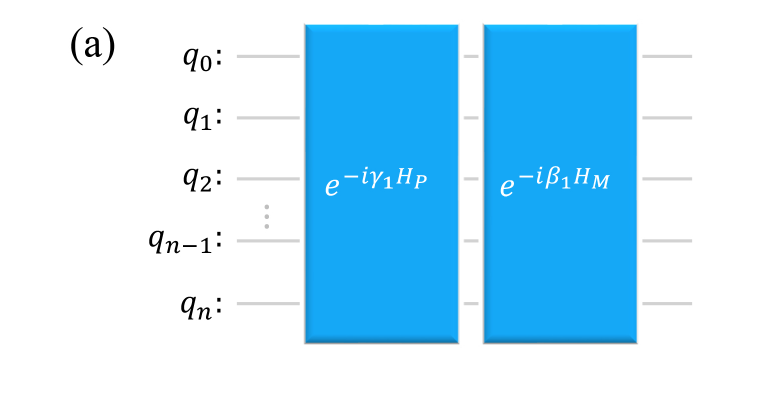}
	\end{minipage}
	\begin{minipage}{0.48\linewidth}
		\centering
		\label{fig:11(b)}\includegraphics[width=\linewidth]{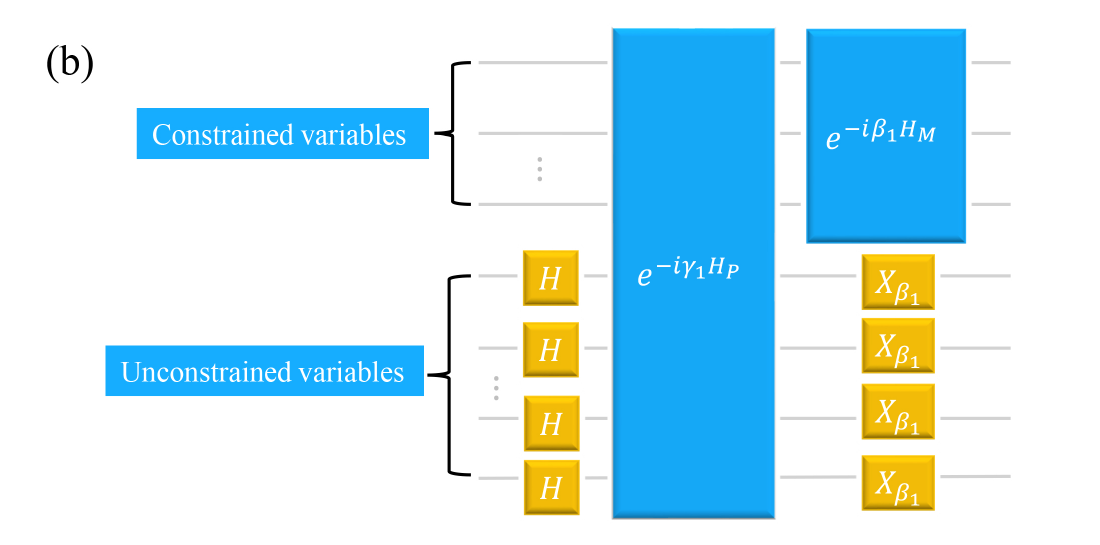}
	\end{minipage}
    \caption{(a) One layer QAOA+ circuit for constrained optimization problems without unconstrained variables. (b) One layer QAOA+ circuit for constrained optimization problems including unconstrained variables, where $X_{\beta_1}$ represents the single-qubit rotating gate $R_X$.}
\label{Fig.11}
\end{figure*}

Variational Quantum Algorithms (VQAs) are a hot class of hybrid quantum-classical algorithms, promising to realize quantum advantages on Noisy Intermediate-Scale Quantum (NISQ) devices \cite{17,18}. The ansatz design of VQAs is important, and ansatz is generally divided into two types: problem-agnostic ansatz and problem-inspired ansatz. The structure of the problem-agnostic ansatz carries no information about the problem itself and is mostly suited for optimization problems, such as the Hardware-Efficient Ansatz (HEA) \cite{Hardware-efficient}, which has the advantage of reducing the circuit depth as much as possible and being able to be implemented efficiently on a quantum chip. The specific structure of the problem-inspired ansatz typically depends on the task at hand, such as Quantum Approximation Optimization Algorithm (QAOA) \cite{QAOA} and Quantum Alternating Operator Ansatz (QAOA+) \cite{31} to solve the combinatorial optimization problems, which have been applied to many problems \cite{maxcut2018,maxcut2021,mvcp2022,cc2022,33,34,35,36,37}. HEA and QAOA need to search for the target solution in the Overall Hilbert Space, which may result in an invalid solution. QAOA+ can restrict the state of the system to the Feasible Space (a subspace of the entire Hilbert space), resulting in zero probability of obtaining invalid solutions, which implies a prominent advantage compared to QAOA and HEA. For constrained optimization problems without unconstrained variables (variables not included in constraints), one-layer QAOA+ circuit is shown in Figure ~\ref{Fig.11} (a), with the mixed operator $e^{-i \beta_1 H_M}$ applied to all qubits. For constrained optimization problems with unconstrained variables, one-layer QAOA+ circuit is shown in Figure ~\ref{Fig.11} (b). For simplicity, we call such problems as Unconstrained-Variables Problems (UVPs). In Figure ~\ref{Fig.11} (b), the mixed operator $e^{-i \beta_1 H_M}$ acts on the constrained variables, while only the single-qubit rotating gates $R_X$ act on the unconstrained variables. The representation capability of this circuit is limited by the lack of two-qubit gates and the sharing of parameters in the $R_X$, which in turn may affect the performance of QAOA+ to solve UVPs. Therefore, QAOA+ is not suitable for such problems. It is particularly important to design a quantum algorithm for solving UVPs efficiently.

\raggedbottom The Uncapacitated Facility Location Problem (UFLP) \cite{Kuehn and Hamburger} is a combinatorial optimization problem, one of the most important NP-hard problems. However, there still exists a research blank in the field of quantum algorithms for addressing this particular issue. The UFLP can be transformed into a UVP in two steps. First, we transform the inequality constraint into equality constraint by introducing slack variables. Second, to facilitate the construction of the mixed Hamiltonian, a penalty function approach is adopted by including a constraint as part of the objective function, thus obtaining a UVP where the feasible space is composed of bit strings with a fixed Hamming weight 1. In this paper, we propose a novel ansatz, taking the UFLP as an example, that performs mixed operators $e^{-i \beta H_M}$ and HEA act on different qubits respectively. This ansatz is called the Variational Quantum Algorithm-Preserving Feasible Space (VQA-PFS) since it preserves the feasible space. To verify the efficiency of the algorithm, we simulate it for UFLP with 12 instances using MindQuantum \cite{51}, showing that it has at least 54\% higher success probability and requires only about 125 iterations to converge to the optimum, with faster and better convergence compared to QAOA+, QAOA, and HEA. In addition, the circuit depth of our algorithm is reduced by at least 75\%, and the number of CNOT gates and parameter gates are reduced by at least 59\% and 53\%, respectively, compared to QAOA+ and QAOA.

This paper is organized as follows. In Sec. II, the VQA-PFS is proposed. In Sec. III, we apply VQA-PFS to solve UFLP. In Sec. IV, numerical results and analysis are given. Finally, the conclusion is given in Sec. V.
\section{VQA-PFS}
\label{Sec:2}
In QAOA+ \cite{31}, the variational ansatz consists of $p$ layers, each containing a phase separator Hamiltonian $H_P$ and a mixer Hamiltonian $H_M$:
\begin{align}
|\psi_p(\overrightarrow{\gamma}, \overrightarrow{\beta})\rangle = U(H_M, \beta_p)U(H_P, \gamma_p) \cdots U(H_P, \gamma_1)|x\rangle,
\end{align}
where $U(H_M, \beta_j) = e^{-i \beta_j H_M}$, $U(H_P, \gamma_j) = e^{-i \gamma_j H_P}$, $j=1, 2,\cdots, p$, and the initial state $|x\rangle$ is a trivial feasible solution, or the uniform superposition state of trivial feasible solutions, and $\overrightarrow{\gamma} = (\gamma_1, \gamma_2, \cdots, \gamma_p)$ and $\overrightarrow{\beta} = (\beta_1, \beta_2, \cdots, \beta_p)$ are variational parameters sets. The variational parameters are optimized on classical computer with the goal of finding the optimal parameters $(\overrightarrow{\gamma^*}, \overrightarrow{\beta^*})$, which are obtained by minimizing the expected value of the phase separator Hamiltonian $H_P$
\begin{align}
  (\overrightarrow{\gamma^*}, \overrightarrow{\beta^*})\ = arg \ \mathop{min}\limits_{\overrightarrow{\gamma}, \overrightarrow{\beta}}\ F_p(\overrightarrow{\gamma}, \overrightarrow{\beta}),
\end{align}
where $F_p(\overrightarrow{\gamma}, \overrightarrow{\beta}) = \langle\psi_p(\overrightarrow{\gamma}, \overrightarrow{\beta})| H_P |\psi_p(\overrightarrow{\gamma}, \overrightarrow{\beta})\rangle.$

\begin{figure*}
 \centering
 \includegraphics[width=0.9\linewidth]{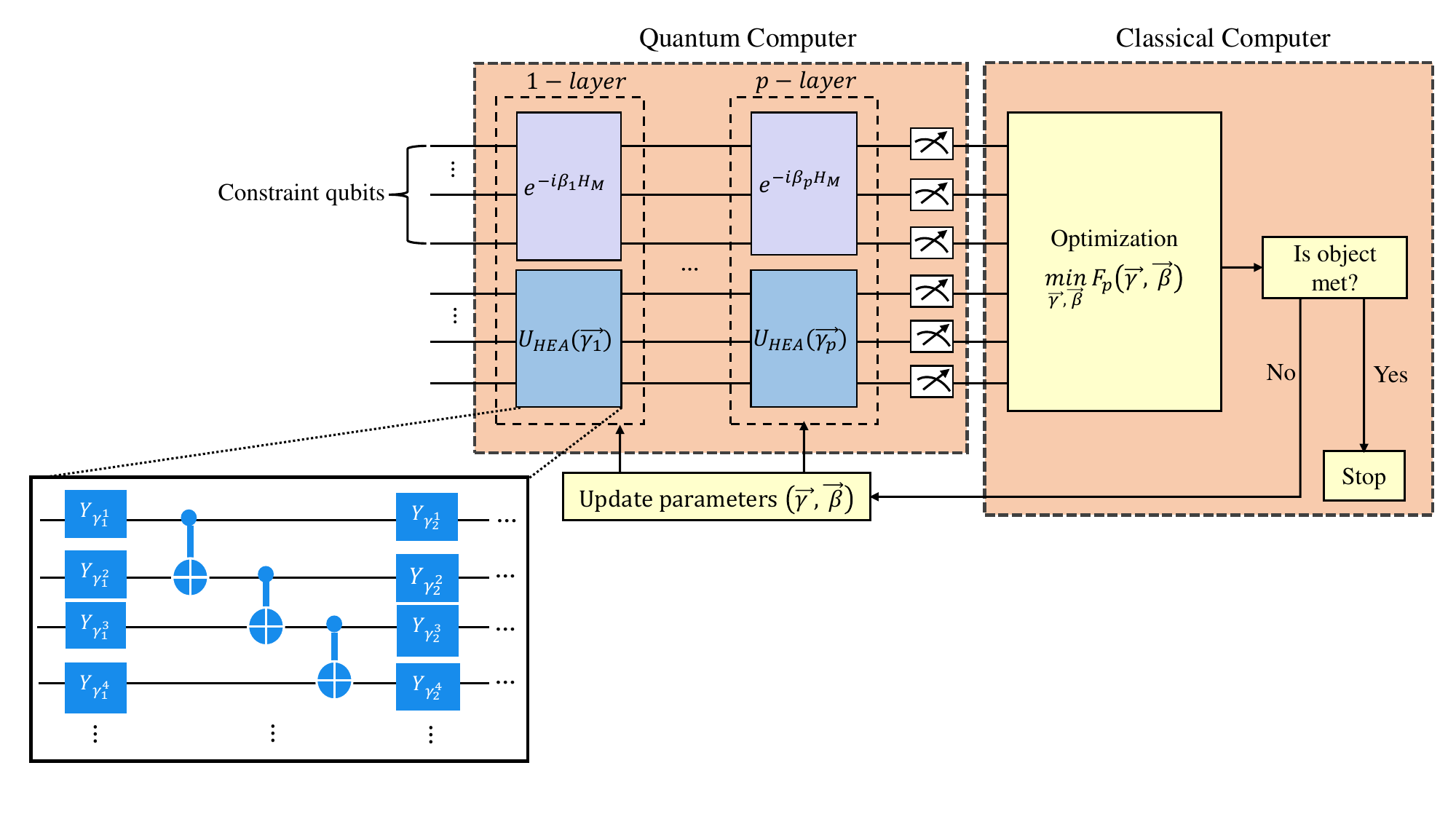}
 \caption{The framework of the VQA-PFS. The variational parameters were optimized on a classical computer, and a quantum computer was used to evaluate the expected value of the objective function. The mixing operator $U(H_M, \beta_i)$ acts on variables of the constraints, and the unitary operator $U_{HEA}(\protect\overrightarrow{\gamma_i})$ of HEA acts on non-constrained variables in the constrained optimization problems.}
 \label{FIG:1}
\end{figure*}

A family of mixing operators $U(H_M, \beta)$ of QAOA+ limits the state of the system to the feasible space, and results in zero probability of obtaining invalid solutions, which is the advantage of the algorithm. HEA has the advantage of reducing the circuit depth as much as possible and being easy to implement efficiently on a quantum chip \cite{Hardware-efficient}. For UVPs, one-layer QAOA+ circuit is shown in Figure ~\ref{Fig.11} (b), with the mixed operator $e^{-i \beta_1 H_M}$ acting on the constrained variables, and only the single-qubit rotating gates $R_X$ acting on the unconstrained variables. The representation capability of this circuit is limited by the lack of two-qubit gates and the sharing of parameters in the $R_X$, which in turn may affect the performance of QAOA+ to solve UVPs. Therefore, QAOA+ is not suitable for such problems. For UVPs, we consider performing mixed operators $U(H_M, \beta)$ on constrained variables and HEA on unconstrained variables. To reduce the circuit depth, the phase-separation operators $U(H_P, \gamma)$ are removed, resulting in the following ansatz
\begin{align}
|\psi_p(\overrightarrow{\gamma}, \overrightarrow{\beta})\rangle = U(H_M, \beta_p)U_{HEA}(\overrightarrow{\gamma_p}) \cdots U_{HEA}(\overrightarrow {\gamma_1})|x\rangle,
\end{align}
where $\overrightarrow {\gamma_i}=(\gamma_i^1, \gamma_i^2, \cdots, \gamma_i^l)$, $i=1,2,\cdots, p$, and $l$ is the number of unconstrained variables. $U(H_M, \beta_i)$ acts on variables of the constraints, and $U_{HEA}(\overrightarrow {\gamma_i})$ acts on unconstrained variables in constrained optimization problems. The crucial points of VQA-PFS are the initial state $|x\rangle$, the mixing operators $U(H_M, \beta)$, the unitary operators $U_{HEA}(\gamma)$, and the phase separator Hamiltonian $H_P$. The framework of the VQA-PFS is shown in Figure ~\ref{FIG:1}.

To evaluate the quality of the solution for the optimization problem, we define the success probability as the probability of finding the optimal solution
\begin{align}
P_{success} = |\langle x_{sol}|\psi_p(\overrightarrow{\gamma}, \overrightarrow{\beta})\rangle|^2,
\end{align}
where $x_{sol}$ is an optimal solution to the problem.

\section{Apply VQA-PFS to solve UFLP}
\label{Sec:3}
 The UFLP is a combinatorial optimization problem that can be transformed into a variant of UVP. In this section, we first introduce the definition and the mathematical model of UFLP. Then, we solve UFLP using VQA-PFS.

\subsection{UFLP}
\label{Subsec:3.1}
The definition of UFLP is generally described as \cite{Zhang2023}: giving the set $K = \{k_1, \cdots , k_m\}$ of customers, where $m$ is the number of customers and $k_i$ is the $i$th customer. Giving the set $S = \{s_1, \cdots , s_n\}$ of potential facilities that can be opened, where $n$ is  the number of facilities and $s_j$ is the $j$th facility. Giving an $m \times n$ matrix $D = \left [d_{ij} \right]_{m \times n}$, where $d_{ij}$ represents the service cost when the $i$th customer receives the service from the $j$th facility. $G = \{g_1, \cdots , g_n\}$ is the set of fixed open cost of facilities, where $g_j$ represents the opening cost required by the opening of the $j$th facility. It is worthy to note that the demand of any customer is fulfill by only one facility. The goal of UFLP is to find a set of open facilities and a reasonable allocation scheme between facilities and customers, so that the sum of total service costs and total open facility costs is minimized.

The mathematical model of UFLP \cite{Zhang2023} is described as follows
\begin{align}
& min    \quad  \sum_{i=1}^m\sum_{j=1}^n d_{ij}y_{ij} + \sum_{j=1}^n g_jx_j,  \label{eq4} \\
& s. t.  \quad  \sum_{j=1}^n y_{ij} = 1, \quad i=1,2,\cdots,m, \label{eq5}\\
& \quad\quad y_{ij}\leq x_j, \quad i=1,2,\cdots,m,  j=1,2,\cdots,n,\label{eq6}\\
& \quad \quad y_{ij}, x_j \in \{0, 1 \}, \label{eq7}
\end{align}
where $y_{ij}=1$ if $i$th customer gets service from $j$th facility; otherwise $y_{ij}=0$, and $x_{j}=1$ if $j$th facility is opened; otherwise $x_{j}=0$.
The number of binary decision variables is $mn +n$. The first term in the Eq. (\ref{eq4}) denotes the total service cost, and the second term denotes the total opening cost of the opened facilities. The Eq. (\ref{eq5}) ensures that every customer is served by exactly one facility. The Eq. (\ref{eq6}) ensures that a customer can be served from a facility only if a facility is opened.

To facilitate solving UFLP, we transformed the above optimization problem into the following standard form by introducing slack variables $z_{ij}$
\begin{align}
& min    \quad  \sum_{i=1}^m\sum_{j=1}^n d_{ij}y_{ij} + \sum_{j=1}^n g_jx_j,  \label{eq8} \\
& s. t.  \quad  \sum_{j=1}^n y_{ij} = 1, \quad i=1,2,\cdots,m, \label{eq9}\\
& \quad\quad y_{ij}+ z_{ij}- x_j=0, \quad i=1,\cdots,m,  j=1,\cdots,n,\label{eq10}\\
& \quad \quad z_{ij}, y_{ij}, x_j \in \{0, 1 \}. \label{eq11}
\end{align}
The number of binary decision variables is $2mn +n$.

VQA-PFS performs mixed operators on constrained variables and HEA on unconstrained variables, respectively, combining the advantages of QAOA+ and HEA to find the objective solution in the feasible space while reducing the depth of the circuits, and one of the cores is the construction of mixed Hamiltonian.

To facilitate the creation of the mixed Hamiltonian, we adopt the penalty function approach by making the constraint Eq. (\ref{eq10}) a part of the objective function Eq. (\ref{eq8}), resulting in the following new optimization problem
\begin{align}
& min    \quad  \sum_{i=1}^m\sum_{j=1}^n d_{ij}y_{ij} + \sum_{j=1}^n g_jx_j + \lambda h(x, y, z),  \label{eq12} \\
& s. t.  \quad  \sum_{j=1}^n y_{ij} = 1, \quad i=1,2,\cdots,m, \label{eq13}\\
& \quad \quad z_{ij}, y_{ij}, x_j \in \{0, 1 \}, \label{eq14}
\end{align}
where $h(x, y, z)=\sum_{i=1}^m\sum_{j=1}^n (y_{ij}+ z_{ij}- x_j)^2$, $\lambda$ is the penalty, determined empirically. The number of constraint variables in Eq. (\ref{eq13}) is $mn$, and the number of non-constrained variables in Eq. (\ref{eq12}) is $mn+n$. Next, we will solve the UFLP via VQA-PFS designed in section ~\ref{Sec:2}.

\subsection{VQA-PFS FOR UFLP}
\label{Subsec:3.2}
The crucial points of VQA-PFS are the initial state $|x\rangle$, the mixing operators $U(H_M, \beta)$, the unitary operators $U_{HEA}(\gamma)$, and the phase separator Hamiltonian $H_P$ for UFLP. For initial state $|x\rangle$, according to Eq. (\ref{eq13}), the state $ \overbrace{|10 \cdots 0_{n-1}}^{n} \overbrace {10\cdots 0_{2n-1}}^{n} \cdots \overbrace{10 \cdots0_{mn-1}}^{n}\overbrace {0 \cdots 0_{2mn+n-1}}^{mn+n}\rangle$ can be obtained, which is a trivial feasible solution.

The mixing operators $U(H_M, \beta) = e^{-i \beta H_M}$ depend on Eq. (\ref{eq13}) and its structure, and its core is to construct $H_M$. To maintain the Hamming weight 1 of the constraint Eq. (\ref{eq13}), the mixing Hamiltonian $H_M$ is expressed as follows \cite{31,33}
\begin{align}
H_M = \sum_{i=0}^{m-1}\sum_{j=0}^{n-2} X_{j+i\cdot n}X_{j+i\cdot n+1}+Y_{j+i\cdot n}Y_{j+i\cdot n+1}, \label{eq15}
\end{align}
where $X$, and $Y$ represents Pauli-$X$ operation, and Pauli-$Y$ operation respectively.

The unitary operators $U_{HEA}(\gamma)$ of HEA consist of single-qubit rotating gates and entangled gates as shown in Figure ~\ref{FIG:1}.
The phase separator Hamiltonian $H_P$ is obtained by replacing binary variables $x$, $y$, $z$ in Eq. (\ref{eq12}) with $\frac{I-Z}{2}$
\begin{align}
H_P &= \sum_{i=1}^m\sum_{j=1}^n d_{ij}\frac{I-Z_{n*(i-1)+j-1}}{2} + \sum_{j=1}^n g_j\frac{I-Z_{m*n+j-1}}{2}  \notag \\
&+ \lambda \sum_{i=1}^m\sum_{j=1}^n (\frac{I-Z_{n*(i-1)+j-1}}{2}+ \frac{I-Z_{n*(m+i)+j-1}}{2} \notag \\
&- \frac{I-Z_{m*n+j-1}}{2})^2, \label{eq17}
\end{align}
where $Z$ represents Pauli-$Z$ operation and the subscript represents the qubit of action. After the four key points are structured, we apply VQA-PFS to solve the UFLP and perform numerical simulation experiments.
\section{Numerical Simulation}
\label{Subsec:4}
In this section, we perform numerical experiments using the MindSpore Quantum \cite{51}. We study 12 instances for three different problem sizes of the UFLP given in Table ~\ref{tab1} to benchmark the performance of VQA-PFS. Details of the instances are given in Appendix~\ref{Sec:Instance Informations}, corresponding to 10, 14, and 22 qubits, respectively. To find the optimal variational parameters, an Adam optimizer \cite{Adam2015} is utilized which is updated with gradients by an adaptive moment estimation algorithm. For the selection of initial parameters, the random initialization method is adopted.

\begin{table}[!ht]
\caption{The instances of UFLP.} \label{tab1}
\begin{tabular}{p{50pt}<{\centering}p{50pt}<{\centering}p{50pt}<{\centering}p{70pt}<{\centering}}
\hline
   $m \times n$ &  Qubit & Instances & Optimal Value  \\
\hline
 \multirow {5}{*}{2 $\times$ 2} & \multirow {5}{*}{10}  & Instance 1 & 16  \\

  &  & Instance 2 &  42 \\

  &  & Instance 3 &  30 \\

  &  & Instance 4 &  39 \\

  &  & Instance 5 &  52 \\
\hline
 \multirow {5}{*}{3 $\times$ 2} & \multirow {5}{*}{14} & Instance 6 & 21  \\
  &  & Instance 7 & 42\\
  &  & Instance 8 & 40\\
  &  & Instance 9 & 35\\
  &  & Instance 10 & 43\\
\hline
 \multirow {2}{*}{5 $\times$ 2}  & \multirow {2}{*}{22} & Instance 11 & 82  \\
  &  & Instance 12 &   95\\
\hline
\end{tabular}
\end{table}

\begin{figure*}[htbp]
	\centering
	\begin{minipage}{0.48\linewidth}
		\centering
		\label{fig:2(a)}\includegraphics[width=\linewidth]{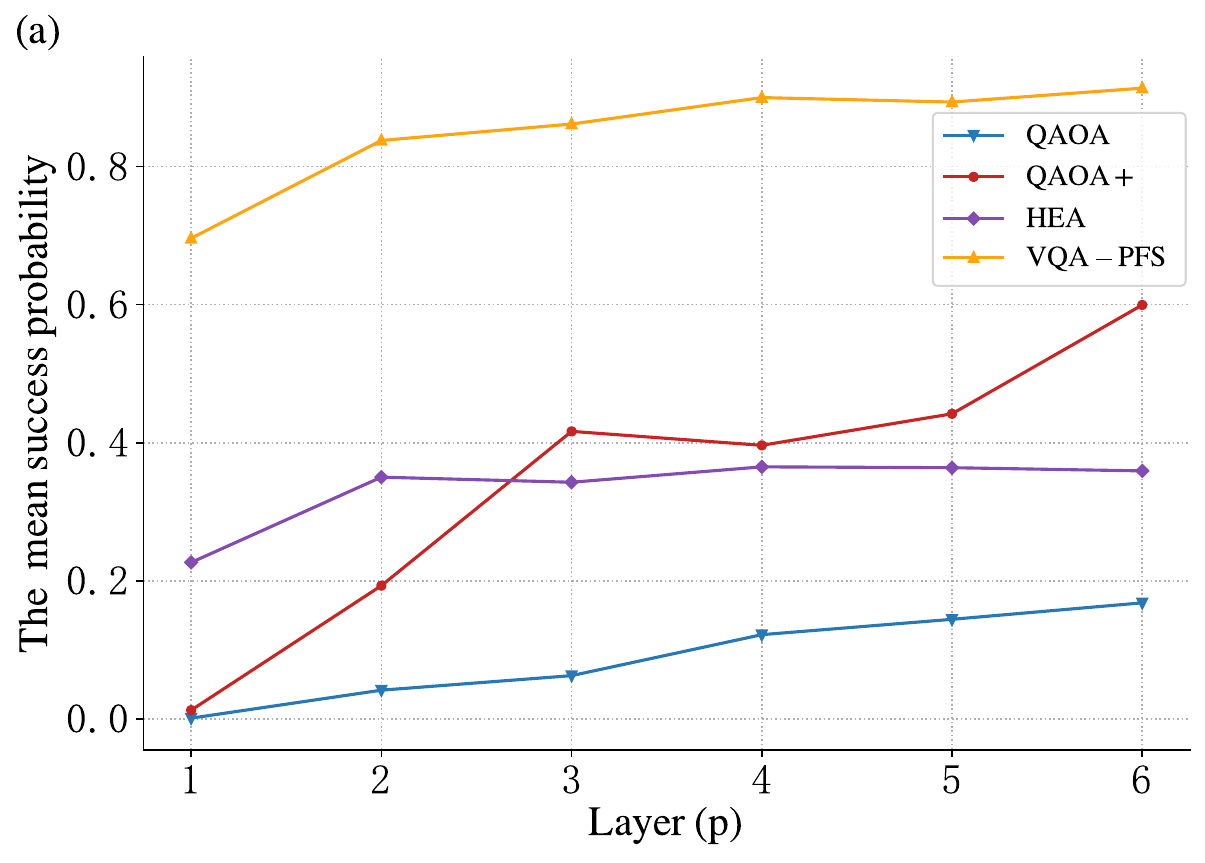}
	\end{minipage}
	\begin{minipage}{0.48\linewidth}
		\centering
		\label{fig:2(b)}\includegraphics[width=\linewidth]{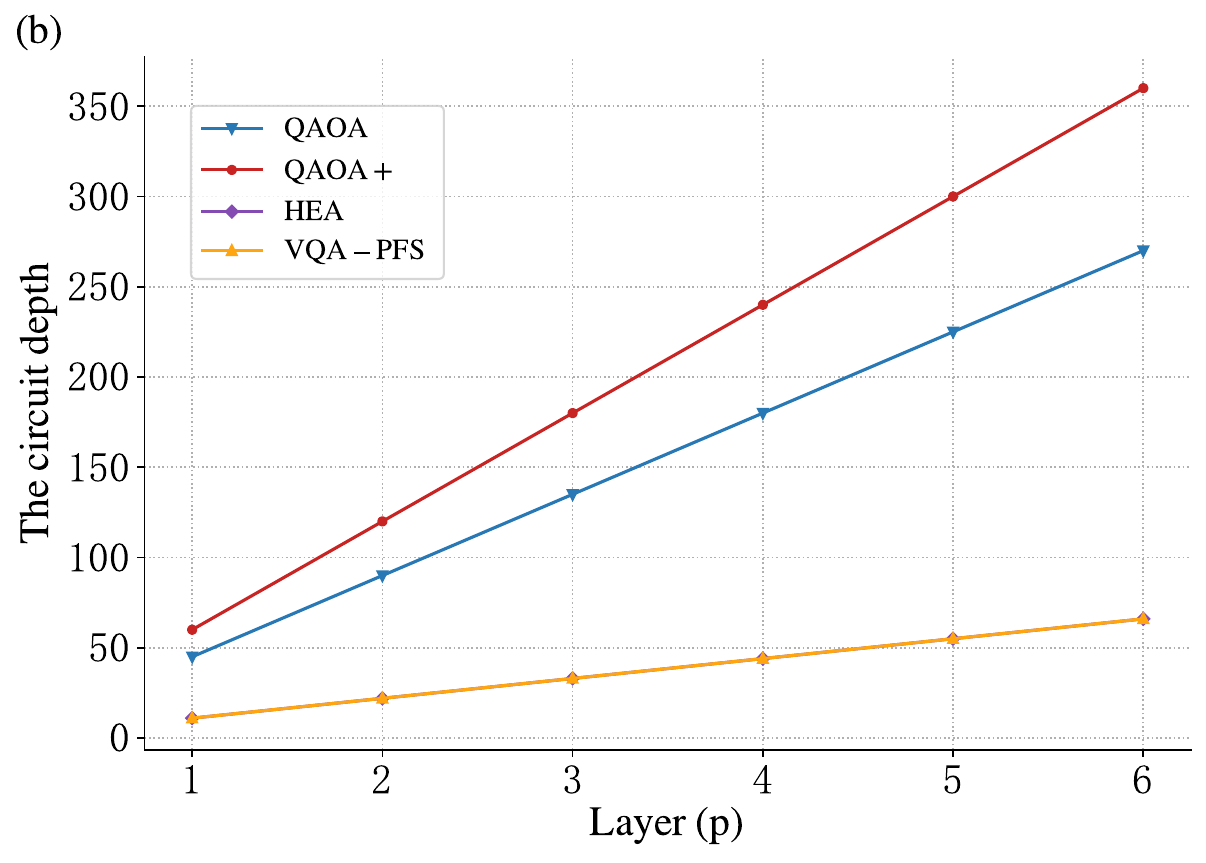}
	\end{minipage}
    \\
    \begin{minipage}{0.48\linewidth}
		\centering
		\label{fig:2(c)}\includegraphics[width=\linewidth]{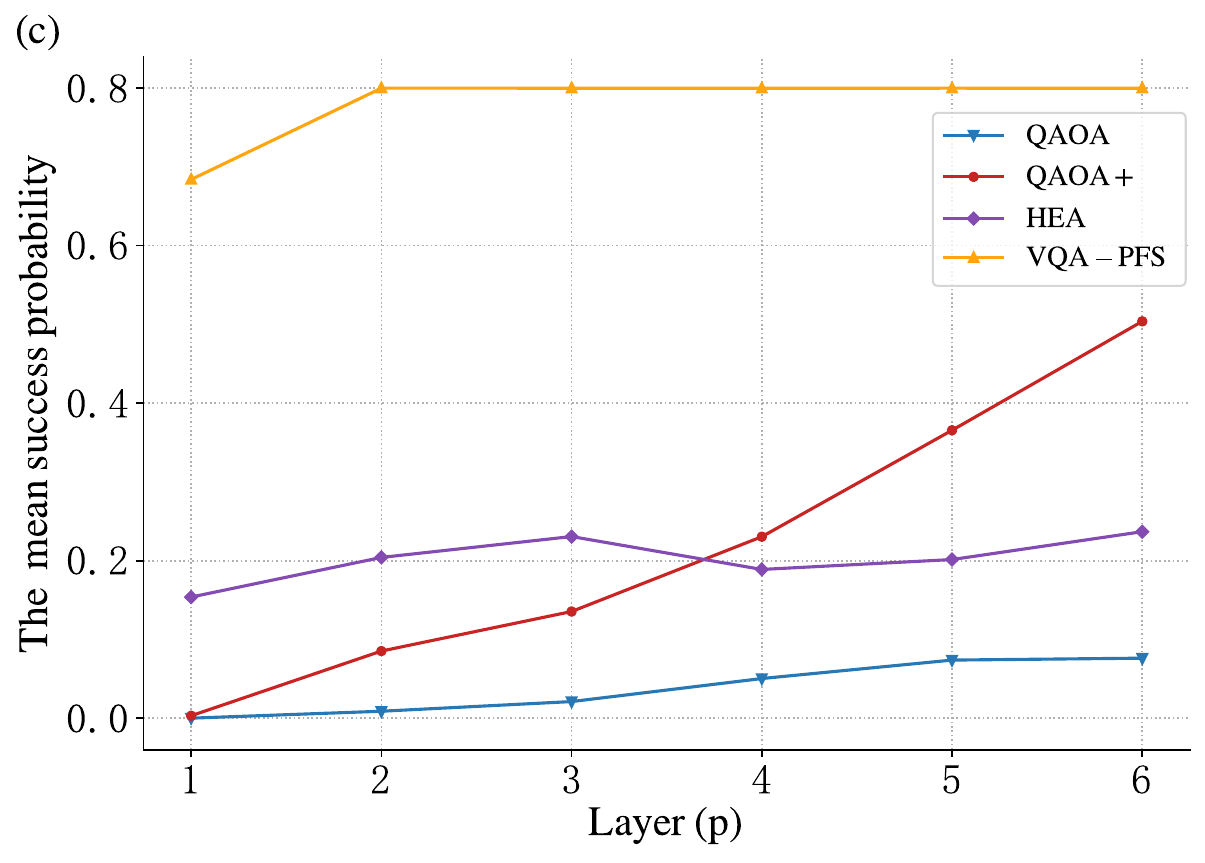}
	\end{minipage}
    \begin{minipage}{0.48\linewidth}
		\centering
		\label{fig:2(d)}\includegraphics[width=\linewidth]{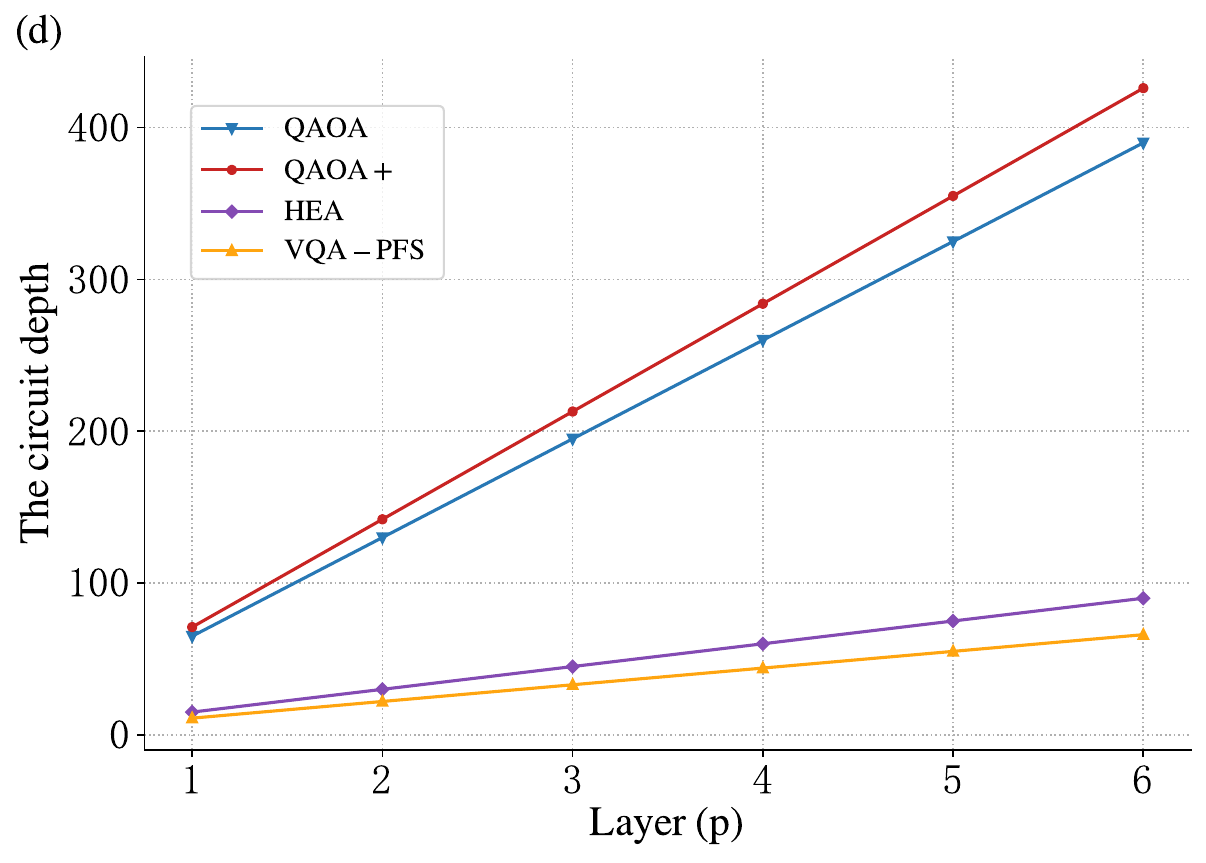}
	\end{minipage}
    \\
    \begin{minipage}{0.48\linewidth}
		\centering
		\label{fig:2(e)}\includegraphics[width=\linewidth]{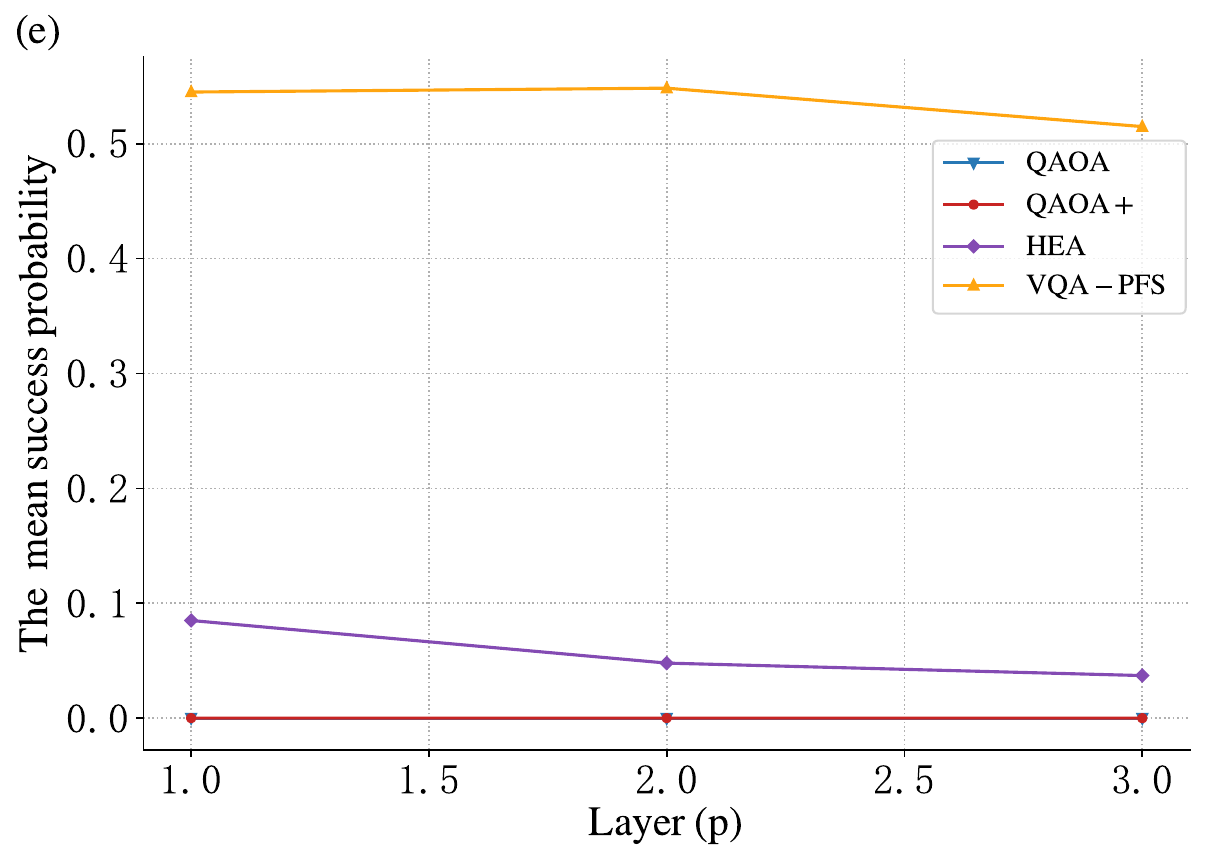}
	\end{minipage}
    \begin{minipage}{0.48\linewidth}
		\centering
		\label{fig:2(f)}\includegraphics[width=\linewidth]{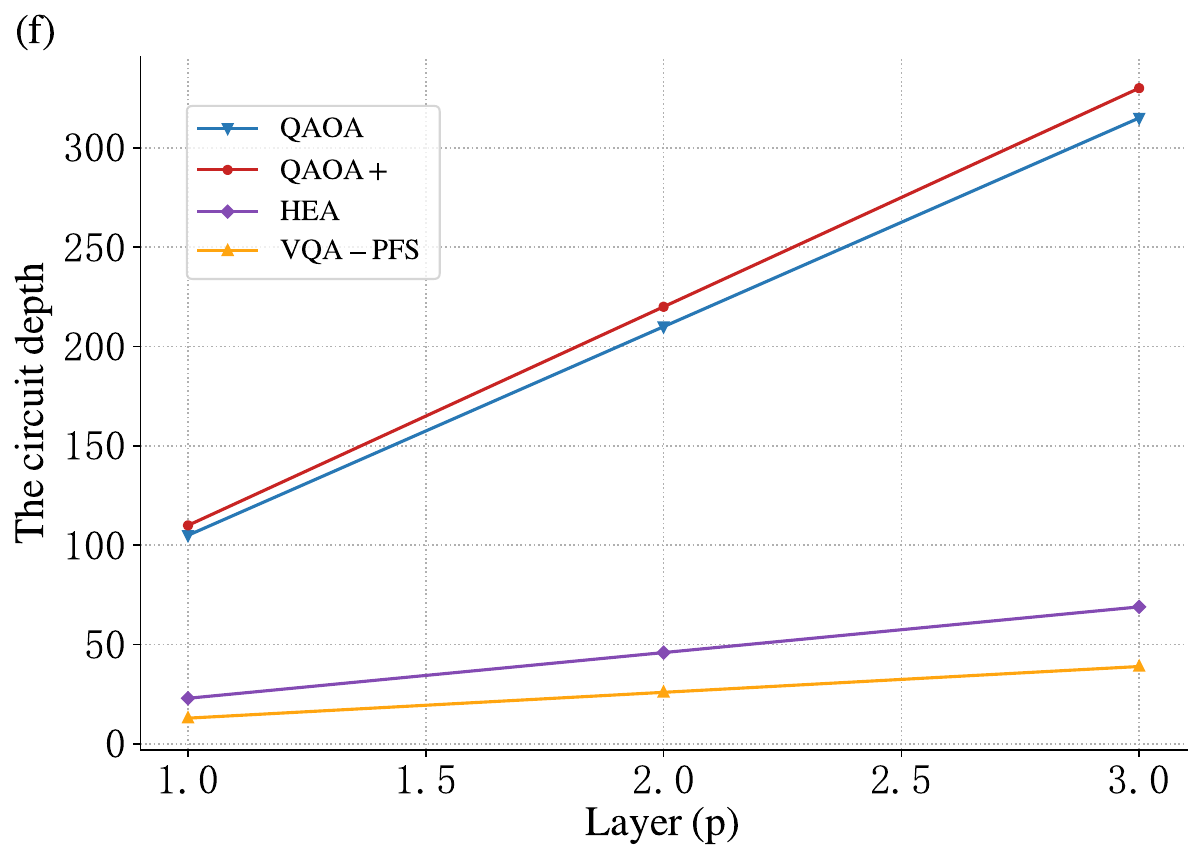}
	\end{minipage}
     \caption{Comparison of the performance of standard QAOA, QAOA+, HEA, and VQA-PFS for the UFLP with 12 instances. The circuit depth and the average success probability as a function of the layers are shown for the standard QAOA, QAOA+, HEA, and VQA-PFS. (a)-(f) The optimization trajectories from optimizer Adam are plotted versus the number of circuit evaluations. For both cases $m=2$, $n=2$ and $m=3$, $n=2$, we perform numerical simulations up to $p = 6$ using random initialization. However, owing to the large number of qubits and high circuit depth involved in case $m=5$, $n=2$, only simulations from $p = 1$ to $p = 3$ are performed. (a)-(b) $m=2$, $n=2$, (c)-(d) $m=3$, $n=2$, (e)-(f) $m=5$, $n=2$.}
\label{Fig.2}
\end{figure*}

To verify the high efficiency of VQA-PFS, we also apply QAOA, QAOA+, and HEA to UFLP and compared them, giving the Hamiltonians and the corresponding circuits, as specified in Appendix~\ref{Sec:QAOA for UFLP}, \ref{Sec:QAOA+ for UFLP}, and \ref{Sec:Hardware Efficient Ansatz for UFLP} respectively. In Fig.~\ref{Fig.2}, we performed numerical simulations up to $p = 6$ using random initialization for 12 instances. The circuit depth and average success probability as a function of the layers are shown for the standard QAOA, QAOA+, HEA, and VQA-PFS, respectively.

Specifically, for $m=2$, $n=2$, in Fig.~\ref{Fig.2} (a) and (b), VQA-PFS has a higher success probability than the other three algorithms, at least 54\% higher. Meanwhile, VQA-PFS has the same circuit depth as HEA, which is at least 75\% lower than QAOA and QAOA+. For $m=3$, $n=2$, in Fig.~\ref{Fig.2} (c) and (d), VQA-PFS has a success probability of at least 58\% higher compared to the other three algorithms. Moreover, the circuit depth of VQA-PFS is about 26\% lower than HEA, and at least 83\% lower than QAOA and QAOA+. For $m=5$, $n=2$, in Fig.~\ref{Fig.2} (e) and (f), compared to the other three algorithms, VQA-PFS has at least five times higher success probability. And, the circuit depth of VQA-PFS is about 43\% lower than HEA, and at least 87\% lower than QAOA and QAOA+. We find that the success probability of VQA-PFS is higher than the other three algorithms by an increasing percentage as the number of qubits increases. Directly, we conjecture that the performance of this algorithm is less affected by the increase in qubits than QAOA, QAOA+ and HEA.
\begin{figure}[htbp]
    \centering
    \includegraphics[width=\linewidth]{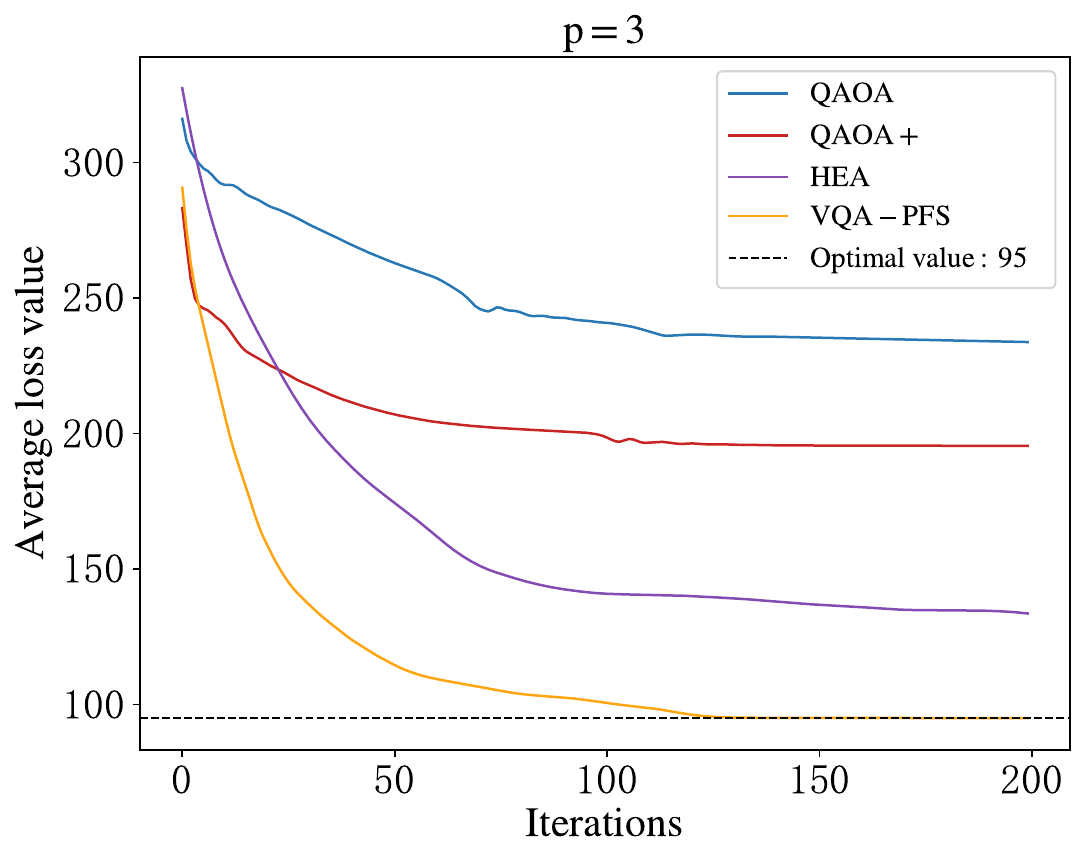}
    \caption{The comparison of the iterations.}
\label{Fig.3}
\end{figure}

\begin{figure}[htbp]
  \centering
  \begin{minipage}[t]{\linewidth}  
      \centering
      \label{fig:4(a)}\includegraphics[width=\linewidth]{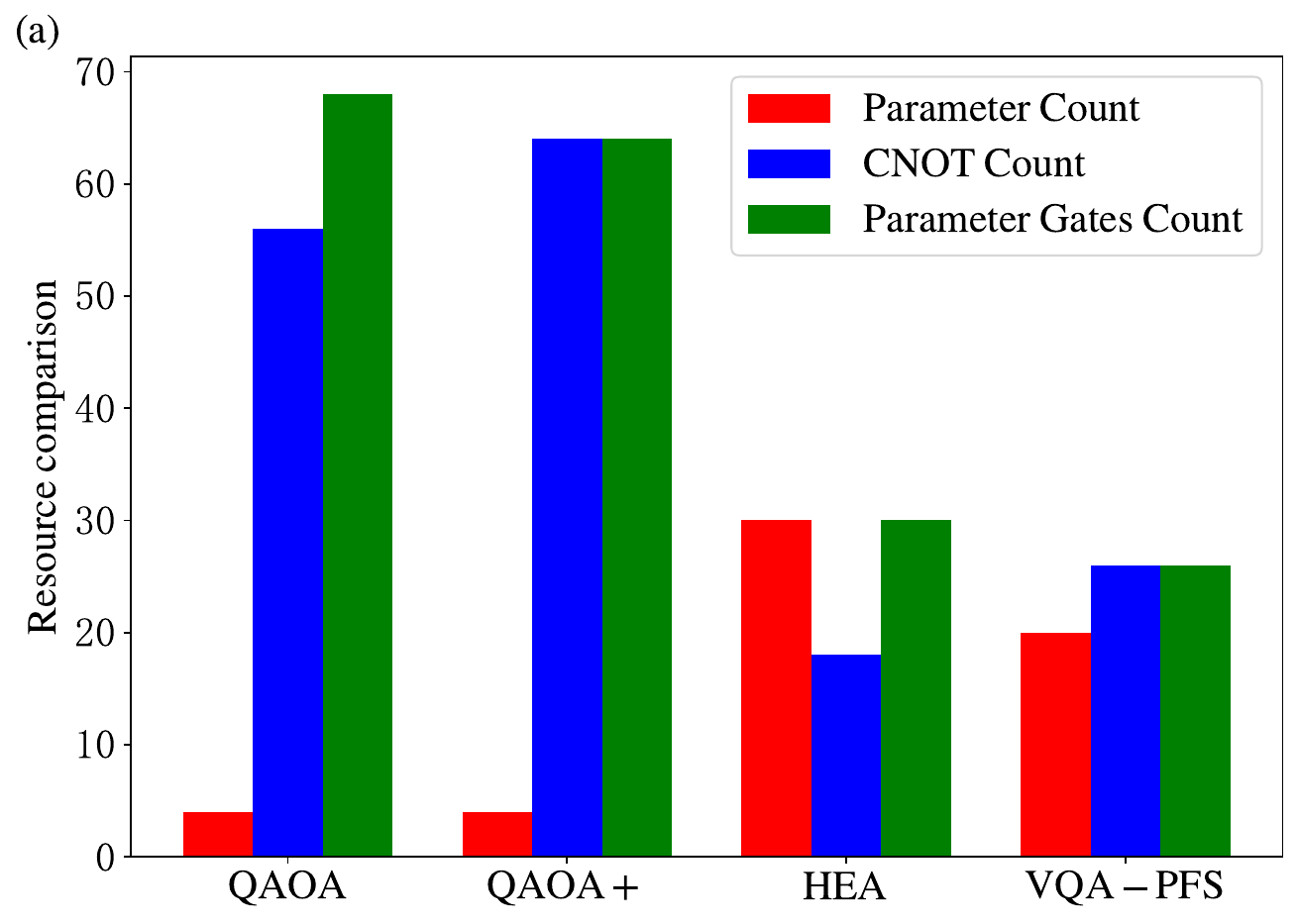}
  \end{minipage}
  \begin{minipage}[t]{\linewidth}
      \centering
      \label{fig:4(b)}\includegraphics[width=\linewidth]{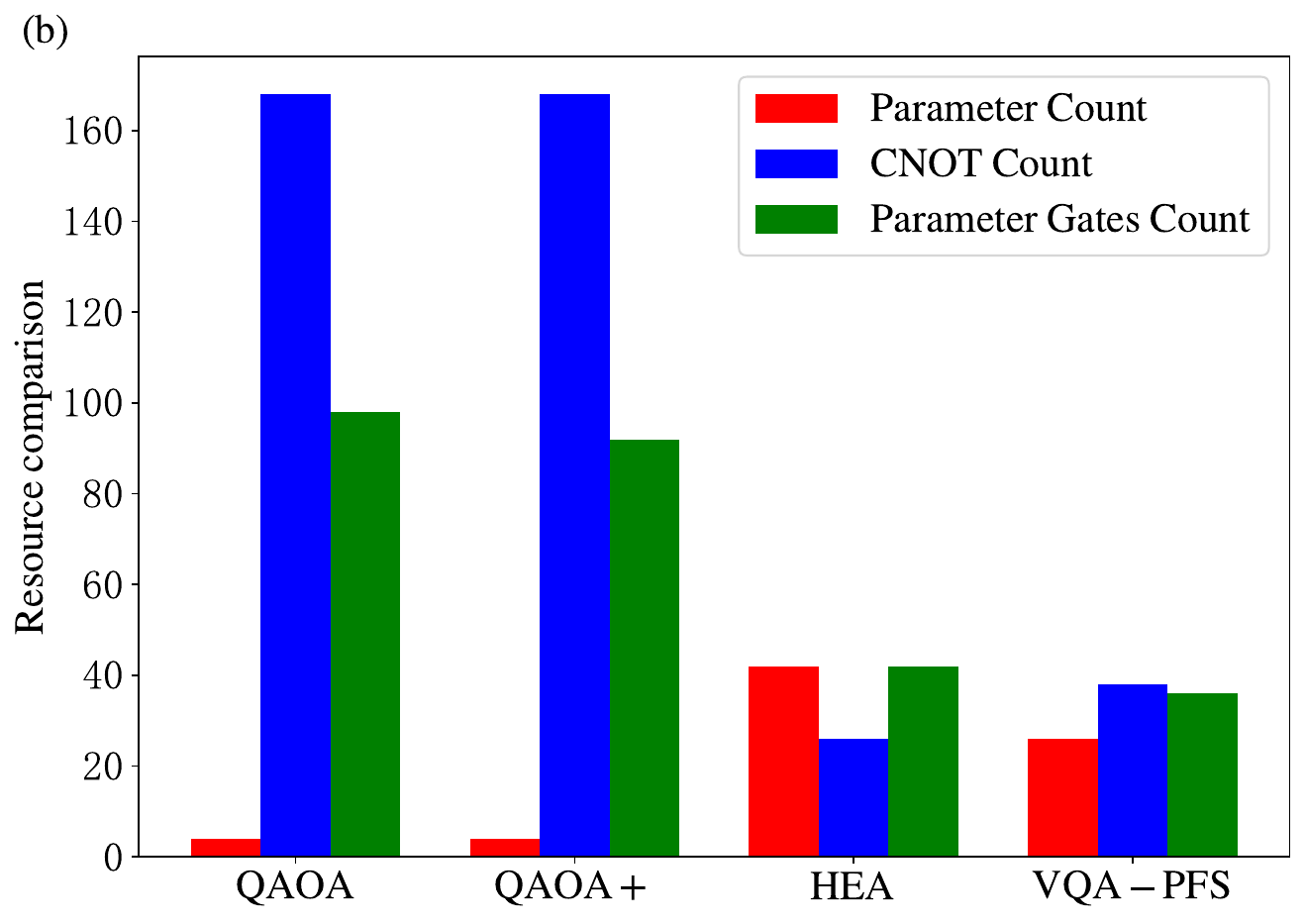}
  \end{minipage}
  \begin{minipage}[t]{\linewidth}
      \centering
      \label{fig:4(c)}\includegraphics[width=\linewidth]{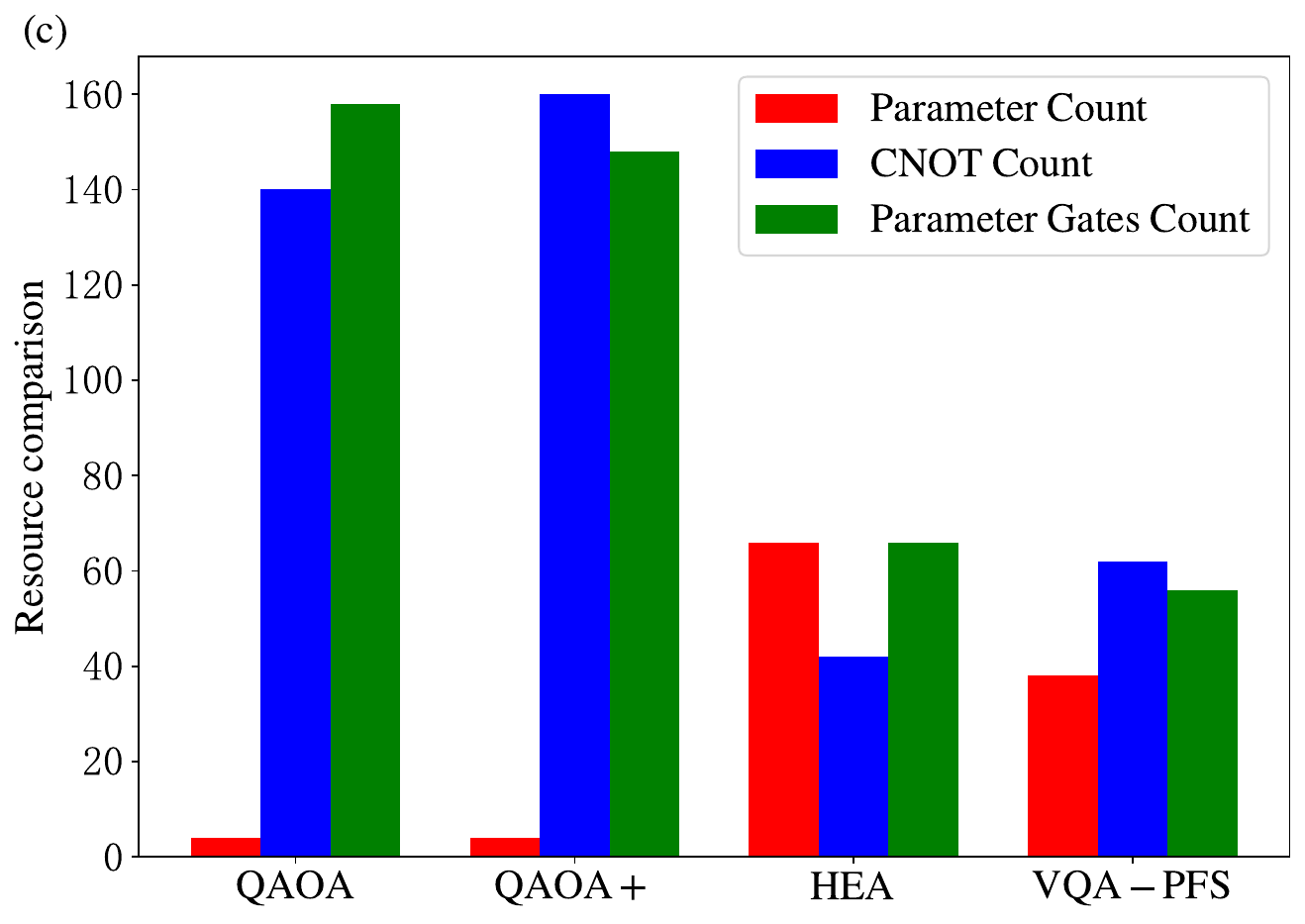}
  \end{minipage}
  \caption{Resource comparison of the standard QAOA, QAOA+, HEA, and VQA-PFS for the UFLP with three different sizes. (a), (b), and (c) show the comparison of four algorithms for the 10, 14, and 22 qubits, respectively.}
  \label{Fig.4}
\end{figure}

The results show that VQA-PFS achieves a significantly higher success probability and lower circuit depth compared to the QAOA, QAOA+, and HEA. One of the keys to these four algorithms is to construct the phase separator Hamiltonian. In QAOA and HEA, for constrained optimization problems, the common method is to incorporate hard constraints into the target function as a penalty item, and then convert the target function into a phase separator Hamiltonian \cite{30}. These two algorithms need to search for the target solutions in the Overall Hilbert Space \cite{31}. Differently, VQA-PFS encodes the constraints directly into quantum circuits, limiting the state of the system to the Feasible Space (a subspace of the entire Hilbert space) of the constrained optimization problems, which implies a higher probability of searching for a solution compared to these two algorithms. Furthermore, for UVPs, the mixed operators $e^{-i \beta H_M}$ of QAOA+ act on the constrained variables, and only the single-qubit rotating gates $R_X$ act on the unconstrained variables (see ~\ref{Sec:QAOA+ for UFLP} for details). The representation capability of this circuit is limited by the lack of two-qubit gates and the sharing of parameters in the $R_X$, which leads to a lower success probability of QAOA+ for solving the UVPs. Different from QAOA+, VQA-PFS removes the phase-separation operators $U(H_P, \gamma)$ and performs the mixed operators $e^{-i \beta H_M}$ and HEA act on different qubits, parallelization can be realized, leading to a lower circuit depth. Therefore, VQA-PFS has lower circuit depth and higher success probability at the same low-layer $p$ compared to the QAOA, QAOA+, and HEA.

To evaluate the convergence speed of our algorithm, we apply QAOA+, QAOA, and HEA to UFLP and compared them. We selected the largest scale instance and plotted Fig.~\ref{Fig.3} through numerical simulation. The loss value as a function of the iterations with $p=3$ is shown for the standard QAOA, QAOA+, HEA, and VQA-PFS, respectively. The results show that our algorithm requires only about 125 iterations to converge to the optimal value, which is faster and better convergence compared to the other three algorithms.

A crucial question, especially for near-term devices, is how the different algorithms compare with respect to resource overhead. We compare quantum resources in terms of the number of CNOT gates, parameter gates, and parameters. Fig.~\ref{Fig.4} shows the number of parameters, CNOTs, and parameter gates for the four algorithms with $p=2$. The results show that our algorithm reduces both the number of CNOT gates by at least 53\% and parameter gates by at least 59\% compared to QAOA and QAOA+, and the number of parameters by at least 33\% and parameter gates by at least 13\% compared to HEA. However, our algorithm has significantly more parameters than QAOA and QAOA+, and more CNOT gates than HEA. In other words, the VQA-PFS obtains better performance in UFLP at the cost of introducing more variational parameters and CNOTs.

\section{CONCLUSION}
\label{Subsec:Dis}
\label{Sec:conclusion}
 In conclusion, we proposed VQA-PFS for solving such UVPs, a combined ansatz that incorporates the advantages of HEA and QAOA+. To verify the high efficiency of VQA-PFS, we also applied QAOA, QAOA+, and HEA to UFLP, as well as compare the performance of these four algorithms. The Hamiltonians and the corresponding circuits are given as specified in Appendix~\ref{Sec:QAOA for UFLP}, \ref{Sec:QAOA+ for UFLP}, and \ref{Sec:Hardware Efficient Ansatz for UFLP} respectively. We tested the performance of several instances of UFLP with 10, 14, and 22 qubits, details of the instances are given in Appendix~\ref{Sec:Instance Informations}, finding that VQA-PFS always outperforms the other three algorithms (see Fig.~\ref{Fig.2} for details).

 Specifically, for $m=2$, $n=2$, in Fig.~\ref{Fig.2} (a) and (b), VQA-PFS has a higher success probability than the other three algorithms, at least 54\% higher. Meanwhile, VQA-PFS has the same circuit depth as HEA, which is at least 75\% lower than QAOA and QAOA+. For $m=3$, $n=2$, in Fig.~\ref{Fig.2} (c) and (d), VQA-PFS has a success probability of at least 58\% higher compared to the other three algorithms. Moreover, the circuit depth of VQA-PFS is about 26\% lower than HEA, and at least 83\% lower than QAOA and QAOA+. For $m=5$, $n=2$, in Fig.~\ref{Fig.2} (e) and (f), compared to the other three algorithms, VQA-PFS has at least five times higher success probability. And, the circuit depth of VQA-PFS is about 43\% lower than HEA, and at least 87\% lower than QAOA and QAOA+. For the largest scale instance, when $p=3$, our algorithm requires only about 125 iterations to converge to the optimal value, which is faster and better convergence compared to the other three algorithms (see Fig.~\ref{Fig.3} for details). In addition, we also compared the number of CNOT gates, parameter gates, and parameters of these four algorithms, and the results showed that our algorithm reduced the number of CNOT gates and parameter gates compared to QAOA and QAOA+, and both parameters and parameter gates are reduced compared to HEA (see Fig.~\ref{Fig.4} for details). Our algorithm is general and instructive for solving such UVPs.
\begin{acknowledgments}
This work is supported by National Natural Science Foundation of China (Grant Nos. 62371069, 62372048, 62272056), Beijing Natural Science Foundation (Grant No. 4222031) and  the 111 Project B21049.
\end{acknowledgments}

\begin{widetext}
\appendix
\section{Details of instances}
\label{Sec:Instance Informations}
In section ~\ref{Subsec:4}, numerical simulations are performed for 12 instances as shown in Table ~\ref{tab2}, Table ~\ref{tab3}, and Table ~\ref{tab4}.
\begin{center}
\begin{table}[!ht]
\caption{The Instance 1-Instance 5 of UFLP.} \label{tab2}
\begin{tabular}{p{50pt}<{\centering} p{50pt}<{\centering} p{100pt}<{\centering} p{180pt}<{\centering} p{70pt}<{\centering} }
\hline
   $m \times n$ &  Qubit & The service matrix $D$ & The set $G$ of fixed open cost of facilities & Optimal Value  \\
\hline \\
 \multirow {13}{*}{2 $\times$ 2} &  \multirow {13}{*}{10}  & $\begin{bmatrix} 6 & 10 \\ 3 & 5 \end{bmatrix}$ & $\left \{ 7, 7 \right \}$ & 16  \\ \\

  &  & $\begin{bmatrix} 16 & 10 \\ 13 & 15 \end{bmatrix}$ & $\left\{17, 17 \right\}$ &  42 \\ \\

  &  & $\begin{bmatrix} 8 & 15 \\ 20 & 15 \end{bmatrix}$ &$\left\{9, 10 \right\}$ &  30 \\ \\

  &  & $\begin{bmatrix} 6 & 20 \\ 13 & 25 \end{bmatrix}$ &$\left\{20, 20 \right\}$ &  39 \\ \\

  &  & $\begin{bmatrix} 25 & 20 \\ 6 & 17 \end{bmatrix}$ &$\left\{27, 15 \right\}$ &  52 \\ \\
\hline
\end{tabular}
\end{table}
\end{center}

\begin{table}[htbp]
\caption{The Instance 6-Instance 10 of UFLP.} \label{tab3}
\begin{tabular}{p{50pt}<{\centering} p{50pt}<{\centering} p{100pt}<{\centering} p{180pt}<{\centering} p{70pt}<{\centering} }
\hline
   $m \times n$ &  Qubit & The service matrix $D$ & The set $G$ of fixed open cost of facilities & Optimal Value  \\
\hline \\
 \multirow {15}{*}{3 $\times$ 2} &  \multirow {15}{*}{14}  & $\begin{bmatrix} 6 & 10 \\ 3 & 1\\ 5 & 4  \end{bmatrix}$ & $\left\{7, 7\right\}$ & 21  \\ \\

  &  & $\begin{bmatrix} 16 & 10 \\ 13 & 5\\ 4 & 10 \end{bmatrix}$ &$\left\{17, 17\right\}$ &  42 \\ \\

  &  & $\begin{bmatrix} 6 & 10 \\ 3 & 5\\ 4 & 1 \end{bmatrix}$ & $\left\{27, 27\right\}$ &  40 \\ \\

  &  & $\begin{bmatrix} 6 & 20 \\ 3 & 15\\ 24 & 1 \end{bmatrix}$ & $\left\{10, 15\right\}$ &  35 \\ \\

  &  & $\begin{bmatrix} 56 & 10 \\ 23 & 5\\ 4 & 18 \end{bmatrix}$ & $\left\{27, 10\right\}$ &  43 \\ \\
\hline
\end{tabular}
\end{table}

\begin{table}[htbp]
\caption{The Instance 11-Instance 12 of UFLP.} \label{tab4}
\begin{tabular}{p{50pt}<{\centering} p{50pt}<{\centering} p{100pt}<{\centering} p{180pt}<{\centering} p{70pt}<{\centering} }
\hline
  $m \times n$ &  Qubit & The service matrix $D$ & The set $G$ of fixed open cost of facilities & Optimal Value \\
\hline \\
 \multirow {7}{*}{5 $\times$ 2} &  \multirow {7}{*}{22}  & $\begin{bmatrix} 16 & 10 \\ 13 & 15 \\ 14 & 10 \\ 15 & 18 \\ 20 & 25\end{bmatrix}$ & $\left\{7, 7\right\}$ & 82  \\ \\

  &  & $\begin{bmatrix} 16 & 10 \\ 13 & 15 \\ 14 & 10 \\ 15 & 18 \\ 20 & 25 \end{bmatrix}$ & $\left\{17, 17\right\}$ &  95 \\ \\
\hline
\end{tabular}
\end{table}
\section{QAOA for UFLP}
\label{Sec:QAOA for UFLP}
The general framework of QAOA is similar to QAOA+ as shown in section ~\ref{Sec:2}, which we will not review here. Similarly, the crucial points of QAOA are the initial state $|x\rangle$, the mixing operators $U(H_M, \beta)$,  and the phase-separation operators $U(H_P, \gamma)$ for UFLP. The initial state is generally chosen as a uniform superposition state, i.e. $|+\rangle^{\bigotimes N}$, $N=2mn+n$. For the mixing operators  $U(H_M, \beta)= e^{-i \beta H_M}$, and its core is to construct $H_M$, where $H_M$ is set to $\sum_{j=0}^{N-1} X_j$. The crucial to the algorithm is the design of the phase separator Hamiltonian, which includes solutions of the UFLP. For the following standard form of the UFLP mathematical model
\begin{align}
& min    \quad  \sum_{i=1}^m\sum_{j=1}^n d_{ij}y_{ij} + \sum_{j=1}^n g_jx_j,  \label{eqB1} \\
& s. t.  \quad  \sum_{j=1}^n y_{ij} = 1, \quad i=1,2,\cdots,m, \label{eqB2}\\
& \quad\quad y_{ij}+ z_{ij}- x_j=0, \quad i=1,\cdots,m,  j=1,\cdots,n,\label{eqB3}\\
& \quad \quad z_{ij}, y_{ij}, x_j \in \{0, 1 \}. \label{eqB4}
\end{align}

To construct phase separator Hamiltonian $H_{P}$, we adopt the common method of introducing the hard constraint as a penalty term into the objective function \cite{26,27,28,29,30}. The following unconstrained optimization problem is obtained
\begin{align}
& min    \quad  \sum_{i=1}^m\sum_{j=1}^n d_{ij}y_{ij} + \sum_{j=1}^n g_jx_j + \lambda (\sum_{i=1}^m(\sum_{j=1}^n y_{ij}- 1)^2 + \sum_{i=1}^m\sum_{j=1}^n (y_{ij}+ z_{ij}- x_j)^2),  \label{eqB5}
\end{align}
where $\lambda$ is the penalty, determined empirically.

The phase separator Hamiltonian $H_P$ is obtained by replacing binary variables $x$, $y$, $z$ in Eq. (\ref{eqB5}) with $\frac{I-Z}{2}$
\begin{align}
H_P &= \sum_{i=1}^m\sum_{j=1}^n d_{ij}\frac{I-Z_{n*(i-1)+j-1}}{2} + \sum_{j=1}^n g_j\frac{I-Z_{m*n+j-1}}{2} \notag \\
&+ \lambda (\sum_{i=1}^m(\sum_{j=1}^n \frac{I-Z_{n*(i-1)+j-1}}{2}- I)^2
+ \sum_{i=1}^m\sum_{j=1}^n (\frac{I-Z_{n*(i-1)+j-1}}{2}+ \frac{I-Z_{n*(m+i)+j-1}}{2}- \frac{1-Z_{m*n+j-1}}{2})^2), \label{eqB6}
\end{align}
where $Z$ represents Pauli-$Z$  operation and the subscript represents the qubit of action.  After the three key points are structured, we apply QAOA to solve the UFLP and perform numerical simulation experiments, and the corresponding 2-layer quantum circuit is shown in Figure ~\ref{Fig.5}.
\begin{figure*}[htp]
\centering
 \includegraphics[width=0.6\linewidth]{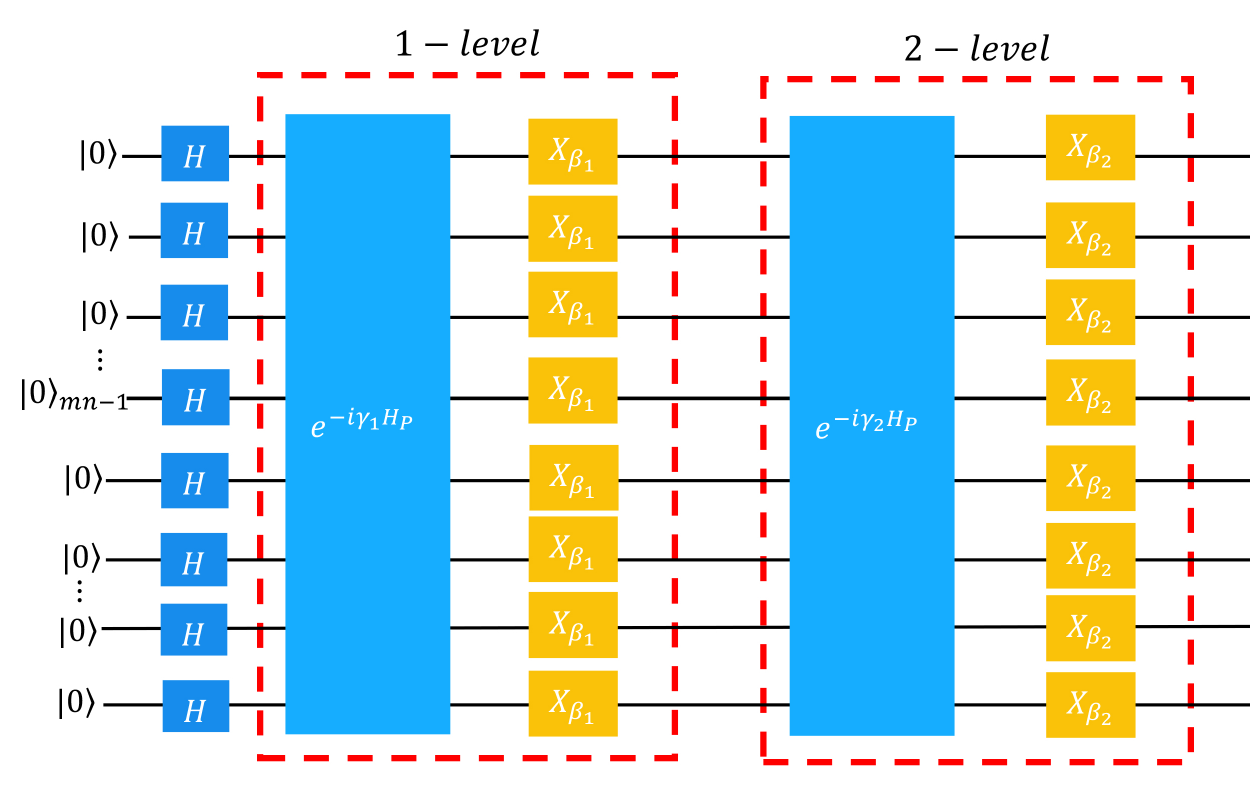}
 \caption{The overall 2-layer circuit of QAOA.}
 \label{Fig.5}
\end{figure*}

\section{QAOA+ for UFLP}
\label{Sec:QAOA+ for UFLP}
The crucial points of QAOA+ are the initial state $|x\rangle$, the mixing operators $U(H_M, \beta)$, the phase-separation operators $U(H_P, \gamma)$ for UFLP.  For initial state $|x\rangle$, according to Eq. (\ref{eq13}), the state $ \overbrace{|10 \cdots 0_{n-1}}^{n} \overbrace {10\cdots 0_{2n-1}}^{n} \cdots \overbrace{10 \cdots0_{mn-1}}^{n}\overbrace {0 \cdots 0_{2mn+n-1}}^{mn+n}\rangle$ can be obtained, which is a trivial feasible solution.

The mixing operators $U(H_M, \beta) = e^{-i \beta H_M}$ depends on Eq. (\ref{eq13}) and its structure, and its core is to construct $H_M$. To maintain the Hamming weight 1 of the constraint Eq. (\ref{eq13}), the mixing Hamiltonian $H_M$ is expressed as follows \cite{31,33}
\begin{align}
H_M = \sum_{i=0}^{m-1}\sum_{j=0}^{n-2} X_{j+i\cdot n}X_{j+i\cdot n+1}+Y_{j+i\cdot n}Y_{j+i\cdot n+1}, \label{C1}
\end{align}
where $X$, and $Y$ represents Pauli-$X$ operation, and Pauli-$Y$ operation respectively.

The phase-separation operators $U(H_P, \gamma)$ depends on Eq. (\ref{eq12}), and its core is to construct $H_P$. The objective function $f = \sum_{i=1}^m\sum_{j=1}^n d_{ij}y_{ij} + \sum_{j=1}^n g_jx_j + \lambda \sum_{i=1}^m\sum_{j=1}^n (y_{ij}+ z_{ij}- x_j)^2$, and the phase separator Hamiltonian $H_P$ is obtained by replacing binary variables $x$, $y$, $z$ in Eq. (\ref{eq12}) with $\frac{I-Z}{2}$
\begin{align}
H_P &= \sum_{i=1}^m\sum_{j=1}^n d_{ij}\frac{I-Z_{n*(i-1)+j-1}}{2} + \sum_{j=1}^n g_j\frac{I-Z_{m*n+j-1}}{2}  \notag \\
&+ \lambda \sum_{i=1}^m\sum_{j=1}^n (\frac{I-Z_{n*(i-1)+j-1}}{2}+ \frac{I-Z_{n*(m+i)+j-1}}{2} - \frac{I-Z_{m*n+j-1}}{2})^2, \label{C3}
\end{align}
where $Z$ represents Pauli-$Z$ operation and the subscript represents the qubit of action. After the three key points are structured, we apply QAOA+ to solve the UFLP and perform numerical simulation experiments, and the corresponding 2-layer quantum circuit is shown in Figure ~\ref{Fig.6}.
\begin{figure*}[htp]
\centering
 \includegraphics[width=0.8\linewidth]{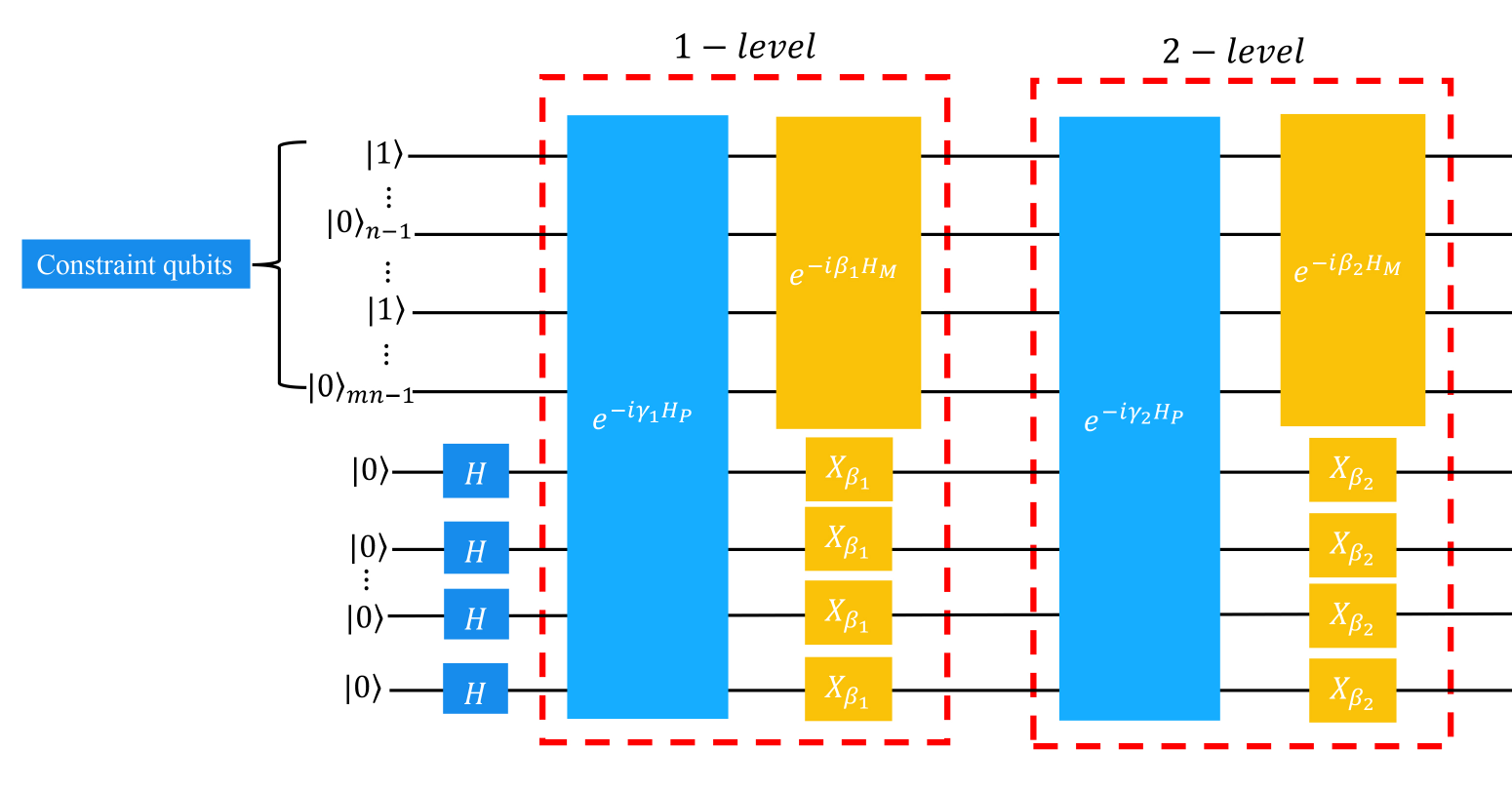}
 \caption{The overall 2-layer circuit of QAOA+.}
 \label{Fig.6}
\end{figure*}

\section{Hardware Efficient Ansatz for UFLP}
\label{Sec:Hardware Efficient Ansatz for UFLP}
The Hardware Efficient Ansatz (HEA) \cite{Hardware-efficient} is a generic name used for ansatzes that are aimed at reducing the circuit depth needed to implement the parametric circuit when using a given quantum hardware. A major advantage of the HEA is its multifunctionality, as it can adapt to encoding symmetries \cite{HEA Bryan, HEA Matthew} and bring the relevant qubits closer to depth reduction \cite{HEA Nikolay}, as well as being particularly useful for investigating Hamiltonians similar to the device{'}s interactions \cite{HEA Christian}.  Optimizing the variational parameters of the circuit is performed on classical computers, aiming to find the optimal parameters by minimizing the expected value of the phase separator Hamiltonian. For UFLP, similar to QAOA, the phase separator Hamiltonian is given in Eq. (\ref{eqB6}). For Instance 1-Instance 5, the corresponding 2-layer quantum circuit is shown in Figure ~\ref{Fig.7}.
\begin{figure*}[htp]
\centering
 \includegraphics[width=0.8\linewidth]{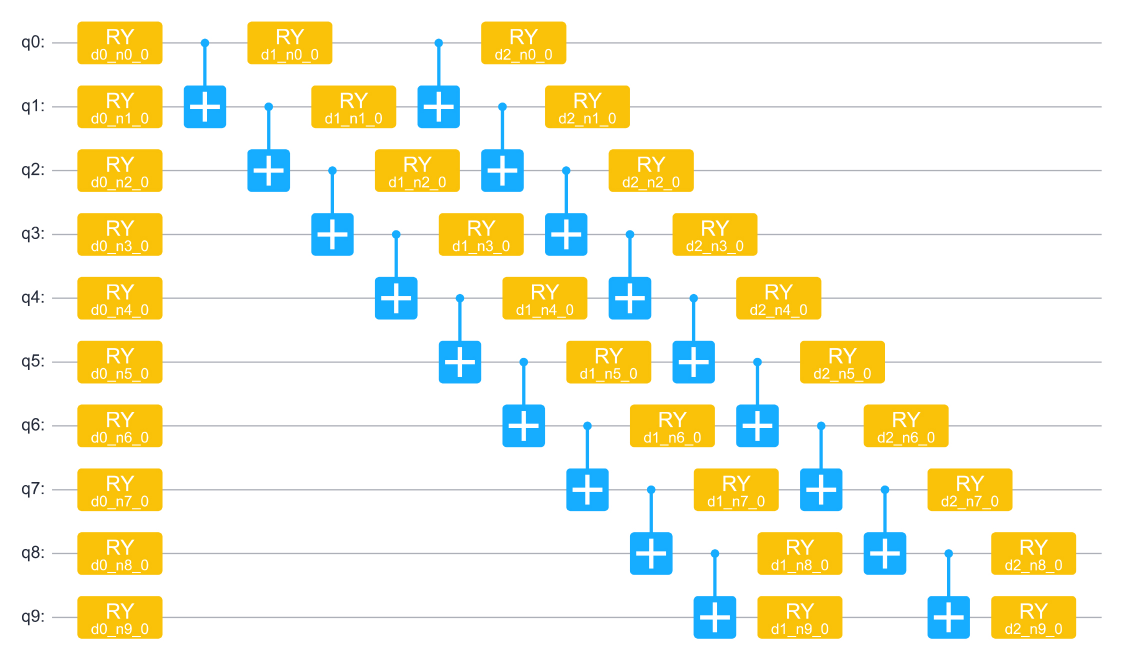}
 \caption{The overall 2-layer quantum circuit of HEA.}
 \label{Fig.7}
\end{figure*}

\end{widetext}
\bibliography{biblatex-phys}

\begin{thebibliography}{11}

\bibitem{1}
P. W. Shor, Algorithms for quantum computation: Discrete logarithms and factoring, in \emph{Proceedings 35th Annual Symposium on Foundations of Computer Science}, pp. 124-134 (1994).
\bibitem{2}
L. K. Grover, A fast quantum mechanical algorithm for database search, in \emph{Proceedings of the Twenty-Eighth Annual ACM Symposium on the Theory of Computing}, pp. 212-219 (1996).


\bibitem{4}
C. H. Yu, F. Gao, C. Liu, D. Huynh, M. Reynolds, and J. Wang, Quantum algorithm for visual tracking, Phys. Rev. A \textbf{99}, 022301 (2019).
\bibitem{5}
C. H. Yu, F. Gao, and Y. Q. Wen, An improved quantum algorithm for ridge regression, IEEE Trans. Know. Data Eng. \textbf{33} (3),  858-866 (2021).



\bibitem{3}
H. Wang, Y. Xue, Y. Qu, et al., Multidimensional Bose quantum error correction based on neural network decoder, npj Quantum Inf \textbf{8}, 134 (2022).

\bibitem{11}
L. C. Wan, C. H. Yu, S. J. Pan, F. Gao, Q. Y. Wen, and S. J. Qin, Asymptotic quantum algorithm for the Toeplitz systems, Phys. Rev. A \textbf{97}, 062322 (2018).
\bibitem{12}
H. L. Liu, L. C. Wan, C. H. Yu, S. J. Pan, S. J. Qin, F. Gao, and Q. Y. Wen, A quantum algorithm for solving eigenproblem of the Laplacian matrix of a fully connected weighted graph, Adv Quantum Technol, 2300031 (2023).
\bibitem{13}
L. C. Wan, C. H. Yu, S. J. Pan, S. J. Qin, F. Gao, and Q. Y. Wen, Block-encoding-based quantum algorithm for linear systems with displacement structures, Phys. Rev. A \textbf{104}, 062414 (2021).
\bibitem{14}
H. L. Liu, C. H. Yu, L. C. Wan, S. J. Qin, F. Gao, and Q. Y. Wen, Quantum mean centering for block-encoding-based quantum algorithm, Physica A: Statistical Mechanics and its Applications \textbf{607}, 128227 (2022).

\bibitem{16}
Z. Q. Li, B. B. Cai, H. W. Sun, H. L. Liu, L. C. Wan, S. J. Qin, Q. Y. Wen, and F. Gao, Novel quantum circuit implementation of Advanced Encryption Standard with low costs, Sci. China-Phys. Mech. Astron. \textbf{65}, 290311 (2022).

\bibitem{17}
J. Preskill, Quantum computing in the NISQ era and beyond, Quantum \textbf{2}, 79 (2018).
\bibitem{18}
J. R. McClean, J. Romero, R. Babbush, and A. Aspuru-Guzik, The theory of variational hybrid quantum-classical algorithms, New J. Phys. \textbf{18}, 023023 (2016).
\bibitem{Hardware-efficient}
A. Kandala, A. Mezzacapo, K. Temme, M. Takita, M. Brink, J. M. Chow, and J. M. Gambetta, Hardware-efficient variational quantum eigensolver for small molecules and quantum magnets, Nature \textbf{549}, 242-246 (2017).
\bibitem{26}
R. Biswas, et al., A NASA perspective on quantum computing: Opportunities and challenges, Parallel Computing \textbf{64}, 81-98 (2017).
\bibitem{27}
E. G. Rieffel, D. Venturelli, B. O'Gorman, M. B. Do, E. M. Prystay, and V. N. Smelyanskiy, A case study in programming a quantum annealer for hard operational planning problems, Quantum Information Processing, \textbf{14} (1), 1-36 (2015).
\bibitem{28}
S. Hadfield, On the representation of Boolean and real functions as Hamiltonians for quantum computing, in \emph{ACM Transactions on Quantum Computing}, pp. 1-21 (2021).
\bibitem{29}
I. Hen and F. M. Spedalieri, Quantum Annealing for Constrained Optimization, Phys. Rev. Applied \textbf{5} (3), 034007 (2016).
\bibitem{30}
V. Choi, Different adiabatic quantum optimization algorithms for the NP-complete exact cover and 3SAT problems, in \emph{Quantum Information $\&$ Computation}, pp. 638-648 (2011).
\bibitem{QAOA}
E. Farhi, J. Goldstone, and S. Gutmann, A quantum approximate optimization algorithm, arXiv:1411.4028.

\bibitem{31}
S. Hadfield, Z. Wang, B. O'Gorman, E. G. Rieffel, D. Venturelli, and R. Biswas, From the quantum approximate optimization algorithm to a quantum alternating operator ansatz, Algorithms \textbf{12} (2), 34 (2019).

\bibitem{maxcut2018}
Z. Wang, S. Hadfield, Z. Jiang, and E. G. Rieffel, Quantum approximate optimization algorithm for maxcut: A fermionic view, Physical Review A \textbf{97}, 022304 (2018).
\bibitem{maxcut2021}
R. Herrman, L. Treffert, J. Ostrowski, P. C. Lotshaw, T. S. Humble, and G. Siopsis, Impact of graph structures for QAOA on maxcut, Quantum Information Processing \textbf{20} (9), 289 (2021).
\bibitem{mvcp2022}
Y. J. Zhang, X. D. Mu, X. W. Liu, X. Y. Wang, X. Zhang, K. Li, T. Y. Wu, D. Zhao, and C. Dong, Applying the quantum approximate optimization algorithm to the minimum vertex cover problem, Applied Soft Computing, \textbf{118}, 108554 (2022).
\bibitem{cc2022}
J. R. Weggemans, A. Urech, A. Rausch, R. Spreeuw, R. Boucherie, F. Schreck, K. Schoutens, J. Min\'{a}\v{r}, and F. Speelman, Solving correlation clustering with QAOA and a Rydberg qudit system: a full-stack approach, Quantum \textbf{6}, 687 (2022).


\bibitem{33}
Z. H. Wang, N. C. Rubin, J. M. Dominy, and E. G. Rieffel, XY mixers: Analytical and numerical results for the quantum alternating operator ansatz, Phys. Rev. A \textbf{10} (1), 012320 (2020).

\bibitem{34}
S. S. Wang, H. L. Liu, S. J.  Qin, F. Gao, and Q. Y. Wen, Quantum Alternating Operator Ansatz for Solving the Minimum Exact Cover Problem, Physica A: Statistical Mechanics and its Applications \textbf{626}, 129089 (2023).
\bibitem{35}
J. Cook, S. Eidenbenz, and A. B\"{a}rtschi, The Quantum Alternating Operator Ansatz on Max-$k$ Vertex Cover, in \emph{APS March Meeting}, \textbf{65}, 1 (2020).
\bibitem{36}
J. Golden, A. B\"{a}rtschi, D. O{'}Malley, and S. Eidenbenz, The Quantum Alternating Operator Ansatz for Satisfiability Problems, in \emph{2023 IEEE International Conference on Quantum Computing and Engineering (QCE)},  pp. 307-312 (2023).
\bibitem{37}
A. Robert, P. K. Barkoutsos, S. Woerner, and I. Tavernelli, Resource-efficient quantum algorithm for protein folding. npj Quantum Inf \textbf{7}, 38 (2021).

\bibitem{Kuehn and Hamburger}
A. A. Kuehn and M. J. Hamburger, A heuristic program for locating warehouses, Management Science, \textbf{9} (4), 643-666 (1963).

\bibitem{Daskin2003}
M. Daskin, L. Snyder, and R. Berger, Facility location in supply chain design, in \emph {Logistics systems: design and optimization}, pp. 39-65 (2005).

\bibitem{Armas and Juan2018}
J. D. Armas and A. A. Juan, A biased-randomized algorithm for the uncapacitated facility location problem, in \emph {Applied mathematics and computational intelligence}, pp. 287-298 (2018).

\bibitem{Cura2010}
T. Cura, A parallel local search approach to solving the uncapacitated warehouse location problem, Computers \& Industrial Engineering, \textbf{59} (4), 1000-1009 (2010).


\bibitem{Lazic2009}
N. Lazic, I. Givoni, P. Aarabi, and B. Frey, FLoSS: Facility location for subspace segmentation, in \emph {IEEE international conference on computer vision}, pp. 825-832 (2009).

\bibitem{Zhang2023}
F. Z. Zhang, Y. C. He, H. B. Ouyang, and W. B. Li, A fast and efficient discrete evolutionary algorithm for the uncapacitated
facility location problem, Expert Systems With Applications \textbf{213}, 118978 (2023).

\bibitem{51}
MindQuantum Developer, MindQuantum, version 0.6.0, https://gitee.com/mindspore/mindquantum (2021).

\bibitem{Adam2015}
D. P. Kingma and J. Ba, Adam: A method for stochastic optimization, in \emph {International Conference on Learning Representations (ICLR)}, pp. 1-15 (2015).
\bibitem{HEA Bryan}
B. T. Gard, L. Zhu, G. S. Barron, N. J. Mayhall, S. E. Economou, and E. Barnes, Efficient symmetry-preserving state preparation circuits for the variational quantum eigensolver algorithm, npj Quantum Information \textbf{6}, 10 (2020).
\bibitem{HEA Matthew}
M. Otten, C. L. Cortes, and S. K. Gray, Noise-resilient quantum dynamics using symmetry-preserving ansatzes, arXiv:1910.06284 (2019).
\bibitem{HEA Nikolay}
N. V. Tkachenko, J. Sud, Y. Zhang, S. Tretiak, P. M. Anisimov, A. T. Arrasmith, P. J. Coles, L. Cincio, and P. A. Dub, Correlation-informed permutation of qubits for reducing ansatz depth in VQE, PRX Quantum \textbf{2}, 020337 (2021).
\bibitem{HEA Christian}
C. Kokail, C. Maier, R. V. Bijnen, T. Brydges, M. K. Joshi, P. Jurcevic, C. A. Muschik, P. Silvi, R. Blatt, C. F. Roos, et al., Self-verifying variational quantum simulation of lattice models, Nature \textbf{569}, 355-360 (2019).
\end{thebibliography}
\end{document}